\documentclass[a4paper,dvips 2t]{revtex4}
\usepackage{axodraw}
\usepackage{setspace}
\usepackage{epsfig,graphics,graphicx,amsmath,amssymb}
\usepackage{rotate}
\usepackage{latexsym}

\begin{document}

\title{\bf\bf One-Loop Radiative  Corrections to the QED Casimir Energy}

 \author{ Reza Moazzemi}\email{r.moazzemi@qom.ac.ir} \author{ Amirhosein Mojavezi}

\affiliation{Department of Physics, University of Qom, Ghadir Blvd., Qom 371614-611, I.R. Iran}

\begin{abstract}

In this paper, we investigate one-loop radiative  corrections  to the  Casimir energy in the presence of  two perfectly conducting  parallel plates for QED theory within the renormalized perturbation theory. In fact, there are three contributions for radiative corrections to the  Casimir energy, up to order $\alpha$. Only the two-loop diagram, which is of order $\alpha$, has been computed by Bordag et. al (1985), approximately. Here, up to this order, we consider corrections due to two one-loop terms, i.e., photonic and fermionic loop corrections resulting from renormalized QED Lagrangian, more precisely. Our results show that only the fermionic loop has a very minor correction and the correction of photonic loop vanishes.
\end{abstract}
\maketitle
\section{Introduction}

 The Casimir effect is a physical manifestation of changes in the zero point energy of a quantum field for different configurations. The zero point configuration refers to one in which there does not exist any on-shell physical excitation of the field.

In 1948 Casimir predicted the existence of this effect as an attractive force between two infinite parallel uncharged perfectly conducting plates in vacuum \cite{Casimir}. This effect was observed experimentally by Sparnaay \cite{Sparnaay} and Arnold et al \cite{arnold} (for a general review on the Casimir effect, see Refs.\cite{Bordag,Miltonn}). Similar measurements have been done for other geometries, and their precisions have been greatly improved \cite{Force,Braden,nostres,a,b,c}. The manifestations of the Casimir effect have been studied in many different areas of physics. For example, the magnitude of the cosmological constant has been estimated using the Casimir effect \cite{d,e,f}. This effect has been also studied within the context of string theory \cite{g}. It has been investigated in connection with the properties of the spacetime with extra dimensions \cite{h,i,j}. The majority of the investigations related to the Casimir effect concerns with the calculation of this energy or the consequence forces for different fields in different geometries, such as parallel plates \cite{Casimir,k}, cubes \cite{l,m,n,o,p,qm,r,s,t}, cylinders \cite{s,u,v,x}, and spherical geometries \cite{s,y,z,cc}.

 Although the Casimir effect has been known for nearly 70 years, the question of what
are the leading radiative corrections to this effect is still a subject  of discussion. The first endeavors  to compute the radiative  corrections to the Casimir energy  were reported in a paper by Bordag, Robaschik, and Wieczorek  (BRW) \cite{hh}. There exist many works on the radiative corrections to the Casimir energy for various cases (see for example \cite{hh,jj,kkk,ll,mm,nn,oo,rr,ss,an,tt}). In the case of a real massive scalar field, Next to Leading Order (NLO) correction to the energy has been computed in \cite{Bordag,t,oo,rr,ss,an,tt,uu,vv,xx,yy,zz}.  Moreover, the two-loop radiative corrections for some effective field theories have been investigated in \cite{kkk,ll,mm}. Bordag and his collaborators have approximately calculated  radiative correction to the Casimir energy due to one of the three related terms of order of $ \alpha $,\raisebox{-3mm}{\includegraphics[scale=1.2]{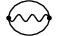}}, in the presence of two perfectly conducting  parallel plates for QED theory. In this viewpoint, the photon propagator satisfies  boundary conditions on the plates, while the plates are transparent to the electrons. They found the correction
$ E ^{(1)} = \frac { \pi^{2} \alpha} {2560ma^{4}}   $
to the  popular leading term of Casimir energy (per unit area) $ E^{(0)}_{\text{em}} = - \frac{\pi^2}{720 a^3}  $, where $a$ is the distance between plates and $ m$ is the electron mass. In 1998 this result with another approach has been reported \cite{jj}. Although they postulate  no  boundary  conditions  for  the  electron
field  because  such  conditions  would  lead  to  additional  contributions  in zeroth order which have  not  been
observed, the fermionic term \cite{Schwinger}, $ E ^{(0)} _{\text{fermion}} = -\frac { m^{2}} {4\pi^2a} \sum\limits_{n=1}-\frac{1}{j^2}[2K_2(2amj)-K_2(4amj)] $  is exist (Here, $K_n(x)$ is the modified Bessel function of order $n$.) However, due to its Yukawa form for the large mass case, $ E ^{(0)} _{\text{fermion}} \sim -\frac { m^{2}} {4\pi^2a} \sqrt{\frac{\pi}{ma}}e^{-2ma}$, at distances much larger than a few Compton wave length of electron it has really  too small value to be observed.

In the context of perturbation theory we need the renormalization to compute loop diagrams. There are two completely equivalent
methods for renormalization; first, bare perturbation theory: working with the bare parameters and relate them to their physical values at the end of calculations, second, renormalized perturbation theory: using counterterms at
the starting point to absorb unphysical part of parameters. Both of them need renormalization conditions to fix the infinities in certain conditions. The differences between two renormalization procedures are purely a matter of
bookkeeping. In the framework of renormalized perturbation theory for QED there are three vacuum bubbles of order of $\alpha$.  Up to now, all the papers on the Casimir effect, that we are aware of, have not been calculated two of those, namely: photonic loop \raisebox{-3.5mm}{\includegraphics[scale=0.5]{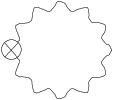}} \ resulting from electromagnetic field  and fermionic loop \raisebox{-3.2mm}{\includegraphics[scale=0.5]{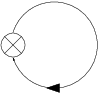}} related to the spinor field. Note that, although according to the common understanding we use the same counterterms for two different situations (with and without plates), the difference of vacuum energies may still be nonzero due to the difference of boundary conditions applying on loop propagators.

The main purpose of  this paper is to directly calculate radiative correction to the Casimir energy resulting from one-loop corrections namely: one-loop photon and one-loop fermion, in the framework of  the renormalized perturbation theory for QED theory.
These corrections are of order $ \alpha$. In order to do this, we use Green's functions in the presence of plates  for Electromagnetic field with Dirichlet boundary condition and for spinor field with MIT bag boundary condition as propagators. Our main regularization is dimensional regularization.

Our approach for the calculation of radiative corrections to Casimir energy is the most direct one. In this way we subtract two infinite energies: one relate to presence  and the other without presence of two plates. We adjust both of their regulators in such a way that the divergences removed and the physical result is obtained.

To have a throughout complete correction, up to order $\alpha$, one must also compute two-loop  term once the fermionic field is
submitted to MIT bag boundary conditions. However, it is notable that almost all the Casimir forces for various massive fields, which precisely calculated in the literature,  have Yukawa asymptotic forms (usually $K_n(ma)$) even for leading term in different dimensions. Therefore, here we only calculate the one-loop diagram which seems more important than two-loop one that has two fermion propagators.

We organized the paper as follows: In Section \ref{secII} we briefly review the renormalization of quantum electrodynamics. In Section \ref{sec3} using analogies between an electromagnetic field and a massless scalar field, photonic loop correction  is considered. We use the Dirichlet boundary condition on the two plates. In Section \ref{sec4} we directly calculate radiative correction to the Casimir energy resulting from fermionic loop where MIT bag boundary condition, as constraints on both of the  plates, is considered. In Section 5 we summarize our results and state our conclusions.

\section{Renormalization of Quantum Electrodynamics: a brief Review}\label{secII}
In this section we  briefly  review systematics of renormalization for QED theory (see for complete details \cite{ABCD}). The original QED Lagrangian is
\begin{equation}\label{11}
{\cal L_{\rm QED}}
=-\frac{1}{4}(F_{\mu \nu})^2 +\bar{\psi} (i\partial \hspace {-2mm}/-m_0)\psi-e_0\bar{\psi}\gamma_{\mu}\psi A^\mu.
\end{equation}
By replacing $\psi= z_2^{\frac{1}{2}}\psi_r$  and $A^{\mu}=z_3^{\frac{1}{2}}A^{\mu}_r $, it becomes
\begin{equation}\label{22}
{\cal L_{\rm QED}}
=-\frac{1}{4}z_3(F^r_{\mu \nu})^2 +z_2 \bar{\psi}_r (i\partial \hspace{-2mm}/ -m_0)\psi_r-e_0z_2z_3^{\frac{1}{2}}\bar{\psi}_r\gamma_\mu\psi_rA_{r}^\mu,
\end{equation}
where $ e_0$ is the bare electric charge and $z_2 $ and $ z_3 $ are the field-strength renormalizations for $ \psi $ and $A^\mu $ respectively.
We define a scaling factor $ z_1 $ as follows:
\begin{equation}\label{33}
ez_1= e_0z_2z_3^{\frac{1}{2}}.
\end{equation}
 We can split each term of the Lagrangian into two pieces as follows:
\begin{equation}\label{Lagrangian}
{\cal L_{\rm QED}}
=-\frac{1}{4}(F^r_{\mu \nu})^2 +\bar{\psi}_r (i\partial\hspace{-2mm}/-m)\psi_r-e\bar{\psi}_r\gamma^\mu\psi_rA_{r}^\mu-\frac{1}{4}\delta_3(F^{\mu\nu}_r)^2
+\bar{\psi}_r(i\delta_2\partial\hspace{-2mm}/-\delta_m)\psi_r
-e\delta_1\bar{\psi}\gamma_\mu\psi_rA_{r}^\mu,
\end{equation}\\
where $\delta_3=z_3-1$, $\delta_2=z_2-1$, $\delta_m=z_2m_0-m$ and $ \delta_1=z_1-1=(\frac{e_0}{e})z_2z_3^{\frac{1}{2}}-1 $ are counterterms. Here, $m$ and $e$ are the physical mass  and physical charge of the electron which measured at large distances.
Now, the Feynman rules for this Lagrangian are
\begin{eqnarray}\label{44}
\begin{picture}(100,50)(0,0)
\CCirc(80,0){2}{Black}{Black}\Photon(80,2)(80,23){2}{2}
\Text(75,20)[b]{{$\mu$}}\Line(60,-12)(80,0)\Line(99,-12)(80,0)
\end{picture}&=&-ie\gamma ^{\mu}
 \\
\begin{picture}(100,50)(0,0)
\CCirc(80,0){5}{Black}{White}\Photon(80,5)(80,25){2}{2}\Line(76.5,-3.5)(83.5,3.5)\Line(76.5,3.5)(83.5,-3.5)
\Text(75,22)[b]{{$\mu$}}\Line(60,-12)(76,-3)\Line(99,-12)(84,-3)
\end{picture}\ &=& -ie\delta_1\gamma ^{\mu}
\\\label{photon}
\begin{picture}(120,40)(0,0)
\Photon(55,0)(105,0){2}{4}\Text(110,-3)[b]{{$\nu$}}
\Text(50,-3)[b]{{$\mu$}}\LongArrow(75,-7)(85,-7)
\Text(80,-17)[b]{{$k$}}
\end{picture} &=& \frac{-i}{k^2+i\epsilon}\bigg(g^{\mu\nu}-(1-\xi)\frac{k^\mu k^\nu}{k^2}\bigg)
\\\label{phver}
\begin{picture}(120,40)(0,0)
\CCirc(80,0){5}{Black}{White}\Photon(55,0)(75,0){2}{2}\Photon(85,0)(105,0){2}{2}
\Line(76.5,-3.5)(83.5,3.5)\Line(76.5,3.5)(83.5,-3.5)\Text(110,-3)[b]{{$\nu$}}
\Text(50,-3)[b]{{$\mu$}}
\end{picture} &=& -i(g^{\mu\nu}k^2-k^{\mu}k^{\nu})\delta_3
\\\label{fermion}
\begin{picture}(120,40)(0,0)
\ArrowLine(55,0)(105,0)\Text(80,-11)[b]{{$p$}}
\end{picture} &=& \frac{i}{p\hspace{-.23cm}\diagup-m+i\epsilon}
\\\label{delta2m}
\begin{picture}(120,40)(0,0)
\CCirc(80,0){5}{Black}{White}\Line(55,0)(75,0)\ArrowLine(85,0)(105,0)\Line(76.5,-3.5)(83.5,3.5)\Line(76.5,3.5)(83.5,-3.5)
\end{picture}&=& i(p\hspace{-.23cm}\diagup\delta_2-\delta_m).
\end{eqnarray}
Each of the four counterterms must be fixed by renormalization conditions. For QED theory these conditions are (see please \cite{ABCD})
\begin{eqnarray}\label{A.2}
\begin{picture}(120,40)(0,0)
\CCirc(80,0){10}{Black}{White}\Line(50,0)(70,0)\ArrowLine(90,0)(110,0)\Text(80,-3)[b]{{1PI}}
\end{picture}&=&-i\Sigma(p)
\\\label{A}
\begin{picture}(120,30)(0,0)
\CCirc(80,0){10}{Black}{White}\Photon(50,0)(70,0){2}{2}\Photon(90,0)(110,0){2}{2}\Text(115,-3)[b]{{$\nu$}}
\Text(45,-3)[b]{{$\mu$}}\Text(80,-3)[b]{{1PI}}
\end{picture}&=&i \Pi^{\mu\nu}(q) =i(g^{\mu\nu}k^2-k^{\mu}k^{\nu})\Pi(k^2)\\\label{A.3}
\left(\begin{picture}(58,35)(50,3)
\CCirc(80,0){10}{Black}{Gray}\Photon(80,10)(80,30){2}{2}
\Text(75,27)[b]{{$\mu$}}\Line(55,-17)(72,-6)\Line(104,-17)(88,-6)
\end{picture}\right)_{\mbox{amputated}}&=&-ie\Gamma^{\mu}(p',p).
\end{eqnarray}

In the above equations $ -i\Sigma(p) $ denotes the sum of all one-particle irreducible (1PI) diagrams with two external fermion lines. By pretending that the photon has a small nonzero
mass $\mu$ to control the infrared divergences, up to leading order in $ \alpha $, the one-loop diagram contributing to $ -i\Sigma(p) $  becomes
\begin{equation}\label{C.0}
 -i\Sigma(p)\mathop {=}\limits_{{\cal O}(\alpha)}  -e^2 \int^1_0dx\int\frac{d^4l}{(2\pi)^4}\frac{-2xp\hspace{-.23cm}\diagup+4m}{[l^2-x(1-x)p^2-x\mu^2-(1-x)m^2]^2}.
\end{equation}
One can evaluate the diagrams in dimensional regularization. If fact, we compute them as an analytic function of the dimensionality of spacetime $d$. The  final expression for any observable quantity should have a well-defined limit as $d\to4$.
Up to leading order in $ \alpha $, $ i\Sigma(p)$ becomes
\begin{equation}\label{C.0}
 -i\Sigma(p)=  -i \frac{e^2m}{(4\pi)^\frac{d}{2}} \int^1_0dx\frac{\Gamma(2-\frac{d}{2})}{\bigg((1-x)^{2}m^{2} + x\mu^2-x(1-x)p^2\bigg)^{2-\frac{d}{2}}}[(4-\epsilon)m-(2-\epsilon)xp].
\end{equation}
with $ \epsilon=4-d $. Since we prefer to work with dimensionless parameters we convert this formula as
\begin{equation}\label{C}
 -i\Sigma(p)=  -i \frac{e^2}{a^{d-3}(4\pi)^\frac{d}{2}} \int^1_0dx\frac{\Gamma(2-\frac{d}{2})}{\bigg((1-x)^{2}\tilde{m}^{2} + x\tilde{\mu}^2-x(1-x)\tilde{p}^2\bigg)^{2-\frac{d}{2}}}[(4-\epsilon)\tilde{m}-(2-\epsilon)x\tilde{p}],
\end{equation}
where $\tilde{l}=la, \tilde{p}=pa, \tilde{\mu}=\mu a,\tilde{m}=ma$. Here $1/a$ is an arbitrary scale with mass dimension 1 (in the problem of Casimir effect $a$ can be the plates separation.)

Moreover, $ i\Pi(k^2)$ defines the sum of all 1PI insertions into the photon propagator and up to order $ \alpha $  becomes
\begin{equation}\label{D}
\Pi(k^2)=\frac{-e^2}{a^{d-4}(4\pi)^\frac{d}{2}}\int^{1}_{0}  dx \frac{\Gamma(2-\frac{d}{2})}{\bigg(\tilde{m}^2-x(1-x)\tilde{k}^2\bigg)^{2-\frac{d}{2}}}8x(1-x),
\end{equation}
where $\tilde{k}=ka$. In Eq.(\ref{A.3}), $ \Gamma^{\mu}(p',p) $ denotes the sum of vertex diagrams. More accurately
\begin{equation}\label{B}
\Gamma^{\mu}(p',p)=\gamma^{\mu} F_1(k^2)+\frac{i\sigma^{\mu\nu}k_{\nu}}{2m}F_2(k^2),
\end{equation}
where $ F_1 $ and $ F_2 $ are unknown functions of $ k^2 $ called form factors and $ \sigma^{\mu\nu}=\frac{i}{2}[\gamma^{\mu},\gamma^{\nu}] $. To lowest order, $ F_1=1 $  and $ F_2=0 $, we have $ \Gamma^{\mu}= \gamma^{\mu}$. By using Eqs. \eqref{D},\eqref{C} and \eqref{B}, up to leading order in $ \alpha $, the counterterms are derived as follows:
\begin{eqnarray}\label{delta3}
  \delta_3&=&\frac{-e^2}{a^{d-4}(4\pi)^\frac{d}{2}}\int^{1}_{0}  dx \frac{\Gamma(2-\frac{d}{2})}{(\tilde{m}^2)^{2-\frac{d}{2}}}8x(1-x),
\\\label{deltam}
 \delta_m&=&\frac{\tilde{m}\delta_2}{a^{d-3}}-\frac{e^2\tilde{m}}{a^{d-3}(4\pi)^\frac{d}{2}} \int^1_0dx\frac{\Gamma(2-\frac{d}{2})}{\left[(1-x)^{2}\tilde{m}^{2} + x\tilde{\mu}^2\right]^{2-\frac{d}{2}}}(4-2x-\epsilon(1-x)),
\\\label{delta2}
  \delta_2&=& \frac{-e^2}{a^{d-4}(4\pi)^\frac{d}{2}} \int^1_0dx\frac{\Gamma(2-\frac{d}{2})}{\left[(1-x)^{2}\tilde{m}^{2} + x\tilde{\mu}^2\right]^{2-\frac{d}{2}}}\left[(2-\epsilon)x-\frac{\epsilon}{2}\frac{2x(1-x)\tilde{m}^2}{(1-x)^{2}\tilde{m}^{2} + x\tilde{\mu}^2}(4-2x-\epsilon(1-x))\right],
\\\nonumber
  \delta_1&=&\frac{-e^2}{a^{d-4}(4\pi)^\frac{d}{2}} \int^1_0dz (1-z)\bigg\{\frac{\Gamma(2-\frac{d}{2})}{((1-z)^{2}\tilde{m}^{2} + z\tilde{\mu}^2)^{2-\frac{d}{2}}}\frac{(2-\epsilon)^2}{2}\\\label{delta1}&&
  \qquad\qquad\qquad\qquad+\frac{\Gamma(3-\frac{d}{2})}{[(1-z)^{2}\tilde{m}^{2} + z\tilde{\mu}^2]^{3-\frac{d}{2}}}[2(1-4z+z^2)-\epsilon(1-z)^2]\tilde{m}^2\bigg\}.
\end{eqnarray}

 According to the above discussion three vacuum bubbles contribute to the Casimir energy: \raisebox{-3.5mm}{\includegraphics[scale=0.5]{g4.eps}}, \raisebox{-3.2mm}{\includegraphics[scale=0.5]{g5.eps}},  \raisebox{-3mm}{\includegraphics[scale=1.2]{4}}. Two first diagrams arise from Eqs. \eqref{phver} and \eqref{delta2m}. Bordag et al. have computed only the last one, though approximately. In the next two sections we will consider the effect of the other two vacuum bubbles.
%\linespread{1}

\section{Photonic Loop}\label{sec3}
In this section, we calculate NLO radiative correction to the Casimir energy due to the photonic loop. We use Dirichlet boundary condition on the two parallel perfectly conducting plates  in (3+1) dimensions. Although electromagnetic field cannot be submitted to Dirichlet boundary conditions itself, one can describe the TE and TM modes of the
electromagnetic field in the presence of the conducting plates as two scalar fields
submitted to Dirichlet boundary conditions.
 Obviously in the presence of the two plates, propagators automatically incorporate the boundary conditions and are position dependent. The  contribution of one-loop photon to the vacuum energy in the interval  $\left(\frac{-a}{2},\frac{a}{2}\right)$  is
\begin{equation}\label{66}
   \Delta E_\text{Ph}
   =\int^{a/2}_{-a/2}
   d^3\mathbf{x}\langle\Omega|{\cal H}_{_I}|\Omega\rangle=i\int^{a/2}_{-a/2}\raisebox{-3mm}{\includegraphics[scale=0.5]{g4.eps}}\quad  d^3\mathbf{x} +{\cal O}(\alpha^2),
\end{equation}
using Eq.\eqref{phver} it becomes
\begin{equation}\label{77}
   \Delta E^{(1)}_\text{Ph}
   =i\int^{a/2}_{-a/2} d^3\mathbf{x} \ D_{B}(x,x) [-i(g^{\mu\nu}k^2-k^{\mu}k^{\nu}) \delta_3],
\end{equation}
where $D_B(x,x^{'})$ is the propagator of electromagnetic field in the bounded space. For overall consistency, we use dimensional regularization to control ultraviolet divergences, and a photon mass $ \mu $ to control infrared divergences. Using analogies between an electromagnetic field and a massless scalar field, photon propagator is considered as
\begin{equation}\label{greenp}
D_B(x,x^{'})
=\frac{-2ig_{\mu\nu}}{a}\int \frac{d\omega}{2\pi}\int\frac{d^{d-2}k_\perp}{(2\pi)^{d-2}}\sum_n
\frac{e^{-i\omega(t-t^{'})}e^{-ik_\perp.(x_\perp-x^{'}_{\perp})}\sin(k_n(z+\frac{a}{2}))\sin(k_n(z^{'}+\frac{a}{2}))}{\omega^2-k_\perp^2-k_n^2+\mu^2}.
\end{equation}
Here $ k_\perp $ and $ k_n  $ denote the  parallel and the perpendicular momenta to plates (in $z$-direction), respectively. Note that, both contributions related to TE mode and TM mode are considered to be the same, hence the final energy should become twice. After the usual Wick rotation and using Eqs. \eqref{delta3} and \eqref{greenp}, with $ x=x' $, and carrying out the integration over the space then over solid angle in the $d$-dimensional Euclidean space we have
\begin{equation}\label{1}
\Delta E^{(1)}_\text{Ph}
=\frac{12S\delta_3 \pi^{\frac{d-1}{2}}}{(2\pi)^{d-1}\Gamma(\frac{d-1}{2})}\int dk_Ek_E^{d-2}\sum_n\frac{k_E^2+k_n^2}{k_E^2+k_n^2+\mu^2},
\end{equation}
where  $S$ is the area of the planes, $ k_E^2 =\omega^2+k_{\perp}^2 $ and $k_n$ is obtained using the Dirichlet boundary condition on the walls,
\begin{equation}
 k_n=\frac{n\pi}{a} ,\quad n= 1, 2, 3, \dots.
 \end{equation}

  The one-loop photon correction to the vacuum energy in free space is
\begin{equation}\label{1010}
   \Delta E'^{(1)}_\text{Ph}
   =i\int d^3\mathbf{x} \ D_F(x,x) [-i(g^{\mu\nu}k^2-k^{\mu}k^{\nu}) \delta_3],
\end{equation}
where $D_F(x, x')$ is the propagator of electromagnetic field in free space in Feynman gauge ($\xi=1$). We can use a trivial periodic boundary condition on the walls located at $-L/2$ and $+L/2$. Carrying out the space integrations gives
\begin{equation}\label{2}
\Delta E'^{(1)}_\text{Ph}
=\frac{12 SL\delta_3\pi^\frac{d-1}{2}}{(2\pi)^d\Gamma(\frac{d-1}{2})}\int dk_Ek_E^{d-2}\int dk\frac{k_E^2+k^2}{k_E^2+k^2+\mu^2}.
\end{equation}
To get a vacuum energy comparable with the volume between plates, we should multiply the above energy by a factor $ \frac{a}{L} $  then take the limit $L\to\infty$. We carry out  $ k_E $ integration and import $\delta_3$ from Eq. (\ref{delta3}). Here we have two types of regulators, $d$ and $\mu$, to control the ultraviolet and infrared divergences, respectively. We first work with $d$ to eliminate some of divergences and derive a result for a photon with mass $\mu$, finally we will approach $\mu$ to zero. As $d\to4$ we
can cancel the divergent terms using our full freedom to choose two
different dimensional regulators $d$ corresponding to free and bounded cases. Then we can perform the integration of $x$ parameter of $\delta_3$ to get
\begin{eqnarray}\label{1212}
\Delta E_\text{Cas.}^\text {Ph}=\Delta E^{(1)}_\text{Ph}-\Delta E'^{(1)}_\text{Ph}+ {\cal O}(\alpha^2)&=&
\frac{2\alpha S\mu'^2}{a^3}\bigg[\sum_{n=1} \sqrt{n^2+\mu'^2}\left(  \gamma-1 +\ln \sqrt{n^{2}+ \mu'^2}\right)\\
&-&\int_{0}^\infty dk' \sqrt{k'^2+\mu'^2}\left(  \gamma-1 +\ln\sqrt{k'^{2}+\mu'^2}\right)
\bigg]+ {\cal O}(\alpha^2),\nonumber
\end{eqnarray}
where $\gamma$ is the Euler-Mascheroni number and we have changed the variables as $k'=\frac{ak}{2\pi}  $ and $\mu'=\frac{a\mu}{2\pi}  $.
Now, we can use the Abel-Plana Summation Formula (APSF)\cite{123}, which basically converts our summation into an integration,
\begin{equation}\label{111}
\sum_{n=1}^\infty f(n)=-\frac{f(0)}{2}+\int_0^\infty f(x) dx +i\int_0^\infty\frac{dt}{e^{2\pi t}-1}[f(it)-f(-it)].
\end{equation}
Apply this formula for Eq.\eqref{1212} yields (see Appendix for details)
\begin{eqnarray}
\Delta E^\text{Ph}_\text{Cas.}&=&
\frac{2S\alpha\mu'^2}{a^3}\Bigg[-\frac{\mu'}{2}\left(  \gamma-1 +\ln \mu'^2\right)\\&&+2\int_{\mu'}^\infty
\frac{dt}{e^{2\pi t}-1} \sqrt{t^2-\mu'^2}\left(  \gamma-1 +\ln \sqrt{t^{2}- \mu'^2}\right)
\Bigg] + {\cal O}(\alpha^2).
\end{eqnarray}
It is obvious that as $ \mu $ tends to zero, $ E^\text{Ph}_\text{Cas.}$ approaches zero, up to order $\alpha$:
\begin{equation}\label{111.1}
\lim_{\mu\to0}\Delta E^\text{Ph}_\text{Cas.}=0.
\end{equation}

Therefore, the photonic loop does not contribute to $ {\cal O}(\alpha) $ radiative correction to the Casimir energy.

\section{ Fermionic Loop}\label{sec4}
 In this section, we calculate NLO radiative correction to the Casimir energy due to fermionic loop \raisebox{-3.2mm}{\includegraphics[scale=0.5]{g5.eps}}. We use the MIT bag boundary condition on the plates. According to  MIT bag boundary condition there is no flux of fermions through the boundary, this means that
\begin{equation}\label{333}
 n_{\mu}j^{\mu}=0,
\end{equation}
 where $ j^\mu$ indicates the current of the Dirac field and $ n_\mu $ is the normal unit vector to the boundary, or more strictly it implies to complete confinement of the spinor field. Note that, ideal conductor boundary condition for the electromagnetic field and bag
boundary conditions for the spinor field  can go together. This can be seen from the field equations (Maxwell equations) written in
the form
\begin{equation}\label{333.33}
 \partial_\mu F^{\mu\nu}=e\bar{\psi}\gamma^\nu\psi,
\end{equation}
 after multiplying with the normal vector $n_\nu$
 \begin{equation}\label{333.34}
 \partial_\mu n_\nu F^{\mu\nu}=e\bar{\psi}n_\nu\gamma^\nu\psi.
\end{equation}
Dirichlet boundary condition on the walls vanishes the left side, so that we can use the bag boundary condition.
 Then MIT bag boundary condition turns out to be \cite{az,ax,ac}
 \begin{equation}\label{444}
  [1+i(\hat{\mathbf{n}}.\boldsymbol{\gamma})]\psi(\mathbf{x})=0,
 \end{equation}
which is satisfied on the boundary, more accurate on the plates. Applying this condition to Dirac spinor field, one can derive
\begin{equation}\label{555}
  pa\cot(pa)=-ma,
\end{equation}
which determines quantized modes. Two limits are interesting to calculate; small mass and large mass limits. As a matter of fact, the mass is
small (large) in comparison with the distance $a$, i.e. $ma\ll1$ ($ma\gg1$). For  small mass limit the solutions of Eq.(\ref{555}) are (see for more details \cite{av})
\begin{equation}\label{eig}
  p_n=(n+\frac{1}{2})\frac{\pi}{a}\ \quad  \text{with }\ n=0,1,2,\dots
 \end{equation}
 where $ p_n $ denotes the parallel momenta  to the plates (in $z$-direction). Now, again for the bounded space we have
\begin{eqnarray}\label{666}
\Delta E_\text{F}
&=&\int^{a/2}_{-a/2}
   d^3x\langle\Omega|{\cal H}_{_I}|\Omega\rangle=i\int^{a/2}_{-a/2} \raisebox{-3mm}{\includegraphics[scale=0.5]{g5.eps}}\quad d^3x + {\cal O}(\alpha^ 2)\nonumber\\&=&i\int^{a/2}_{-a/2} d^3x \ \mbox{Tr}[ S_{B}(x,x) i(p\hspace{-.23cm}\diagup\delta_2-\delta_m)] + {\cal O}(\alpha^ 2),
\end{eqnarray}
where $S_B(x,x^{'})$, the Feynman propagator of spinor field between plates, is
\begin{equation}\label{greenf.0}
S_B(x,x^{'})
=\frac{i}{a}\int\frac{d\omega}{2\pi}\int\frac{d^{2}p_\perp}{(2\pi)^{2}}\sum_{n=0}\frac{p\hspace{-.23cm}\diagup+m}
{\omega^2-p_\perp^2-p_n^2-m^2+i\epsilon}e^{-i\omega(t-t^{'})}e^{-ip_\perp(x_\perp-x{'}_{\perp})}e^{-ip_n(z-z^{'})}.
\end{equation}
Here $ p_\perp $ and $ p_n  $ indicate the  parallel and the  perpendicular momenta to the plates, respectively. Converting the integrals into dimensionless form in $d$ spacetime dimensions we have
\begin{equation}\label{greenf}
S_B(x,x^{'})
=\frac{i}{a^{d-1}}\int\frac{d\tilde{\omega}}{2\pi}\int\frac{d^{d-2}\tilde{p}_\perp}{(2\pi)^{d-2}}\sum_{n=0}\frac{\tilde{p}\hspace{-.23cm}\diagup+\tilde{m}}
{\tilde{\omega}^2-\tilde{p}_\perp^2-p_n^2-\tilde{m}^2+i\tilde{\epsilon}}e^{-i\frac{\tilde{\omega}}a(t-t^{'})}
e^{-i\frac{\tilde{p}_\perp}a(x_\perp-x{'}_{\perp})}e^{-i\frac{\tilde{p}_n}a(z-z^{'})}.
\end{equation}
After the usual Wick rotation and carrying out the integration, one can obtain
\begin{equation}\label{3}
\Delta E_\text{F}^{(1)}
= \frac{32S}{(2\pi)^{d-1}a^{d-1}} \frac{\pi^{\frac{d-1}{2}}}{\Gamma(\frac{d-1}{2})}\int d\tilde{p}^{}_E\tilde{p}_E^{d-2}
\sum_{n=0}\left(\frac{\tilde{p}_E^2+\tilde{p}_n^2 }{\tilde{p}_E^2+\tilde{p}_n^2+\tilde{m}^2}\delta_2-\frac{\tilde{m}}{\tilde{p}_E^2+\tilde{p}_\perp^2+\tilde{p}_n^2+\tilde{m}^2}\delta m\right),
\end{equation}
where $ \tilde{p}_E^2 =\tilde{\omega}^2+\tilde{p}_{\perp}^2 $. Similarly, for the free space we have
\begin{equation}\label{888}
   \Delta E_\text{F}^{'(1)}
    =i\int d^3x \ \mbox{Tr}[S_F(x,x) i(p\hspace{-.23cm}\diagup\delta_2-\delta_m)].
\end{equation}
Using Eq.\eqref{fermion} for the free propagator and after integration we obtain
\begin{equation}\label{4}
\Delta E_\text{F}^{'(1)}=\frac{32 S}{(2\pi)^{d-1}a^{d-2}}\frac{\pi^{\frac{d-1}{2}}}{\Gamma(\frac{d-1}{2})}
\int d\tilde{p}^{}_{E} \tilde{p}_E^{d-2}\int\frac{ d\tilde{p}}{2 \pi}\left(\frac{\tilde{p}_E^2+\tilde{p}^2}{\tilde{p}_E^2+\tilde{p}^2+\tilde{m}^2}\delta_2
-\frac{\tilde{m}}{\tilde{p}_E^2+\tilde{p}^2+\tilde{m}^2}\delta_m\right).
\end{equation}
For the small mass case the radiative correction to Casimir energy corresponding to the fermionic loop becomes
\begin{eqnarray}\label{101010}
\Delta E^\text{F}_\text{Cas.}&=&\nonumber
\Delta E^{(1)}_{\text{F}}-\Delta E'^{(1)}_{\text{F}}+ {\cal O}(\alpha^2)\\
&=&\frac{32 S\pi^{\frac{d-1}{2}}}{(2\pi)^{d-1}\Gamma(\frac{d-1}{2})}\frac{\pi^{d-1}}{a^{d-1}}\int dp_E' p_E'^{d-2}\bigg[\delta_2\bigg(\sum_{n=0}\frac{(n+\frac{1}{2})^2+ p_E'^2}{(n+\frac{1}{2})^2+ p_E'^2+m'}-
\frac{1}2\int dp'\frac{p'^{2}+p_E'^2}{p'^{2}+ p_E'^2+m'^2}\bigg)\nonumber\\
&&\qquad\qquad-\frac{am'\delta_m}{\pi}\bigg(\sum_{n=0}\frac{1}{(n+\frac{1}{2})^2+p_E'^2+m'^2}-\frac{1}2\int dp'\frac{1}{p'^{2}+ p_E'^2+m'^2}\bigg)\bigg] + {\cal O}(\alpha ^2),
\end{eqnarray}
where we use the change of variables as $ p'=\tilde{p}/\pi $, $ p'_E=\tilde{p}_E/\pi$ and $ m'=\tilde{m}/\pi $. Integrating of $ p'_E $ yields

\begin{eqnarray}
\nonumber\Delta E^\text{F}_\text{Cas.}&=&\frac{32 S\pi^{\frac{d-1}{2}}}{2^{d-1}a^{d-1}\Gamma(\frac{d-1}{2})}\frac{1}{2}\sec\left(\frac{d\pi}2\right)\pi\Bigg\{\delta_2m'^2\bigg[\sum_{n=0}^\infty [(n+1/2)^2+m'^2]^{\frac{d-3}{2}}-\int_0^\infty[p'^2+m'^2]^{\frac{d-3}{2}}dp'\bigg]\\\label{11111}
&+&\frac{am'\delta_m}{\pi}\bigg[\sum_{n=0}^\infty [(n+1/2)^2+m'^2]^{\frac{d-3}{2}}-\int_0^\infty[p'^2+m'^2]^{\frac{d-3}{2}}dp'\bigg]\Bigg\}+ {\cal O}(\alpha^2).
\end{eqnarray}
Here we need another type of APSF to convert the sum into integral,
\begin{equation}\label{22222}
\sum_{n=0}^\infty f(n+\frac{1}{2})=\int_0^\infty  f(x)dx-i\int_0^\infty\frac{dt}{e^{2\pi t}+1}[f(it)-f(-it)].
\end{equation}
We can use the following formula to calculate the branch cut integral: if $f(z)=\left(z^n+\alpha^m)\right)^{p/2}$
\begin{equation}\label{22222.2}
 i\int_{0}^{\infty}\frac{f(it)-f(-it)}{e^{2\pi t}+1}dt
   =-2\sin\left(\frac{pn\pi}4\right)\int_{\alpha^{m/n}}^{\infty}
   \frac{\left(t^n-\alpha^m\right)^{p/2}}{e^{2\pi t}+1}dt.
\end{equation}
In addition we know that
\begin{equation}
\frac{1}{e^{2 \pi t}+1}=\sum_{j=1}^\infty (-1)^{j+1} e ^{ -2\pi j t}.
\end{equation}

We use these  formulae, and import $\delta_m$ and $\delta_2$ from Eqs. (\ref{deltam}) and (\ref{delta2}), respectively, into  Eq.\eqref{11111}.
Again, similar to the procedure adopted in photonic loop, which lad to Eq. (\ref{1212}), we first expand the expression about $d=4$ then take the limit $d\to4$. No divergent term remains due to the usual subtraction in Casimir effect. We then, do the $x$ integration. Finally, taking the limit $\mu\to0$ we obtain
\begin{equation}\label{final}
\Delta E^\text{F+Ph}_\text{Cas.}=\Delta E^\text{F}_\text{Cas.}+\Delta E^\text{Ph}_\text{Cas.}
=\sum_{j=1}^\infty- \frac{(-1)^j5S\alpha m^2}{16 j^2\pi^4a } \bigg \{K_{0} (2 jam)+2jam K_1 (2j a m)\bigg[\frac{14}5\ln(ma)+N_j\bigg]\bigg\},
\end{equation}
where $N_j=\gamma-\ln(j\pi^2)-\frac{12+\ln8}5$. This is the final result of the radiative correction to the Casimir energy due to fermionic loop, for the small mass case.

\begin{figure}[th]
\centerline{\includegraphics[width=12.5cm]{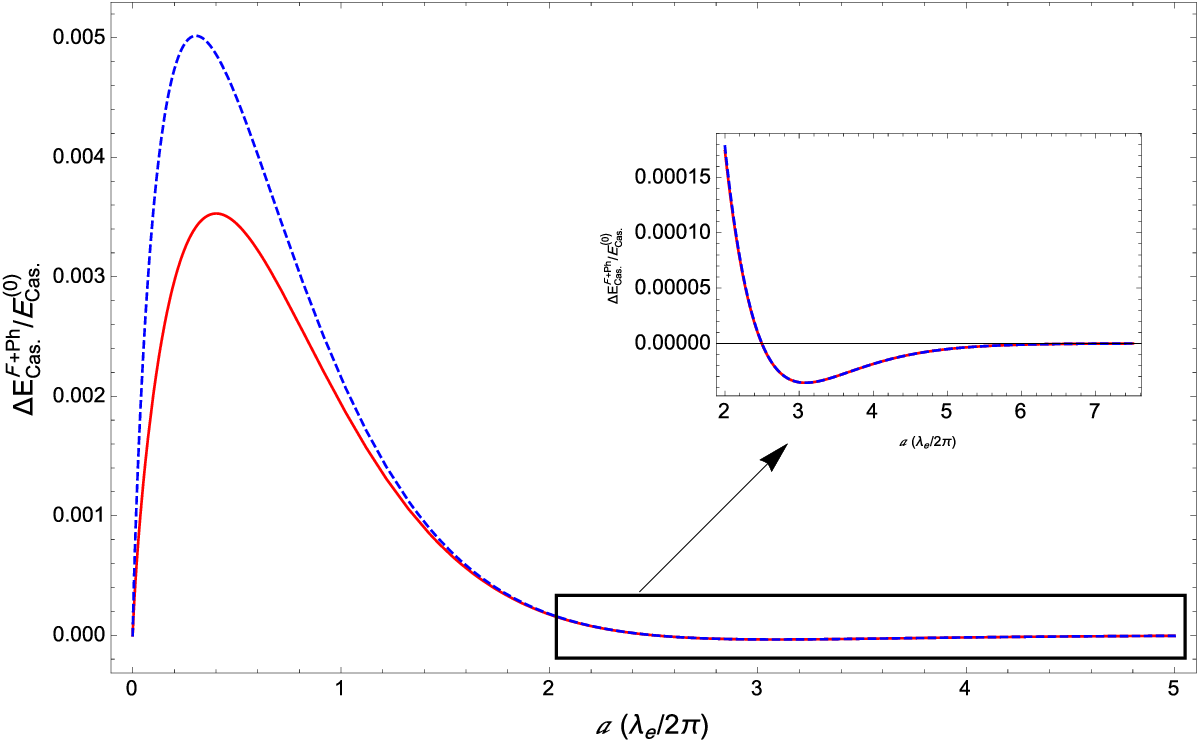}}
\caption{\small  The ratio between the one-loop correction derived here and the leading term of electromagnetic Casimir energy$ E^\text{F+Ph}_\text{Cas.}/E^{(0)}_\text{Cas.}$, vs the plates separation ($\lambda_e$ denotes the Compton wavelength of electron.) Solid (dashed) line shows the large (small) mass limit.
 }\label{fig1}
\end{figure}

\begin{figure}[th]
\centerline{\includegraphics[width=12.5cm]{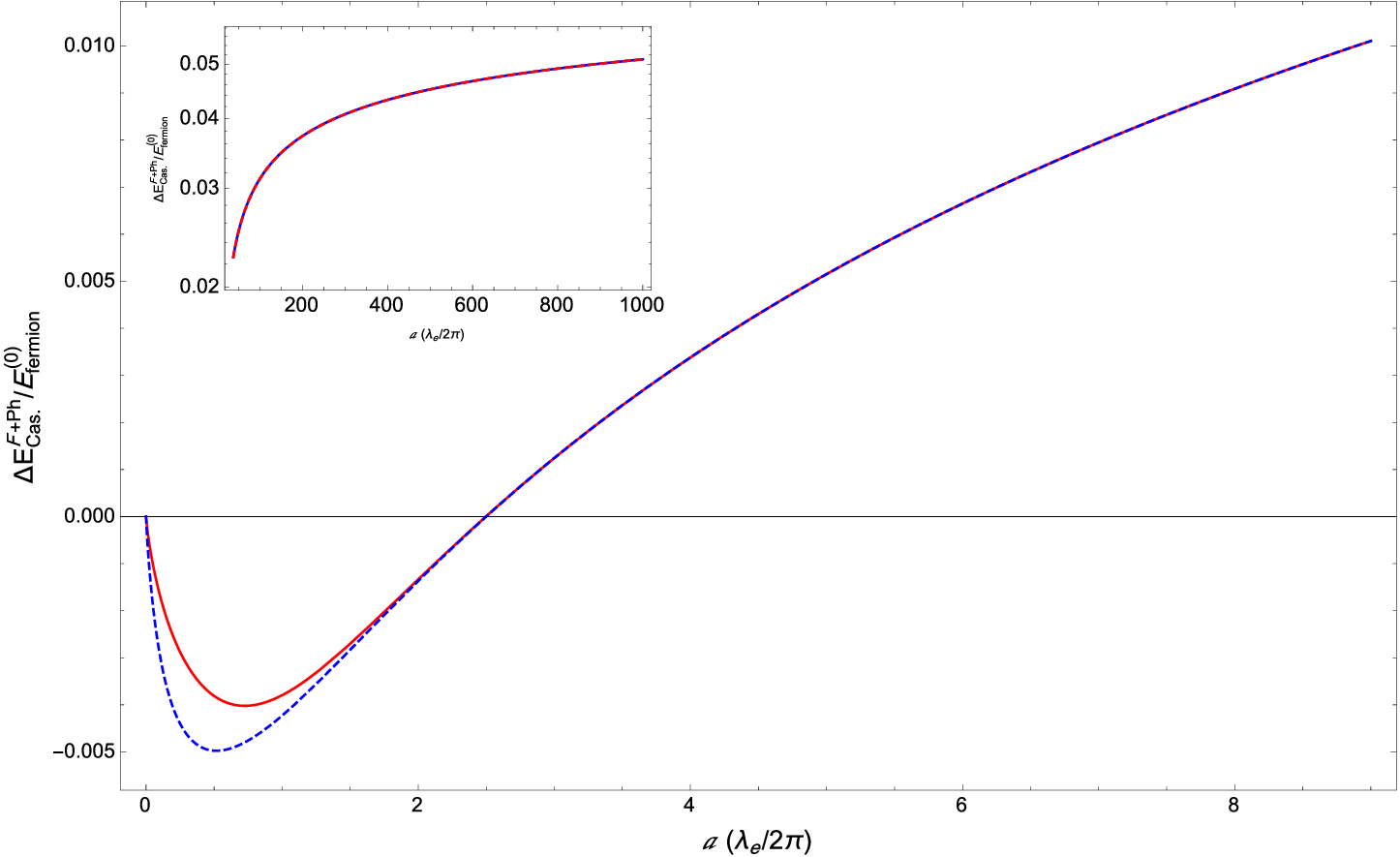}}
\caption{\small  The ratio between the one-loop correction derived here and the leading term of fermionic Casimir energy$ E^\text{F+Ph}_\text{Cas.}/E^{(0)}_\text{Fermion}$, vs the plates separation ($\lambda_e$ denotes the Compton wavelength of electron.) Solid (dashed) line shows the large (small) mass limit.
 }\label{fig2}
\end{figure}

 The other interesting limit is the large mass limit. In this case, Eq.(\ref{555}) turns out to be
 \begin{equation}\label{555.1}
  ma\tan(pa)=-pa.
\end{equation}
Now, the solutions are
 \begin{equation}\label{eig2}
  p_n=\frac{n\pi}{a}\  \text{with }\ n=1,2,\dots.
 \end{equation}
We follow the similar way for obtaining Eq. \eqref{11111}, but now we should apply APSF Eq. (\ref{111}) and need the following relation
\begin{equation}
\frac{1}{e^{2 \pi t}-1}=\sum_{j=1}^\infty e ^{ -2\pi j t}.
\end{equation}
Finally,  our the radiative NLO correction to Casimir energy in this case becomes
\begin{equation}\label{final2}
\Delta E^\text{F+Ph}_\text{Cas.}=\Delta E^\text{F}_\text{Cas.}+\Delta E^\text{Ph}_\text{Cas.}
=\sum_{j=1}^\infty \frac{5S\alpha m^2}{16 j^2\pi^4a } \bigg \{K_{0} (2 jam)+2jam K_1 (2j a m)\bigg[\frac{14}5\ln(ma)+N_j\bigg]\bigg\}
\end{equation}
The first term, in Eq. (\ref{111}) (i.e.  $+f(0)/2$) turns out to be independent of distance between plates $a$. Therefore this term has no impact on the physics of problem and we ignore it. For the large mass case which is also equivalent to the large distances, Eq. (\ref{final2}) takes the form
\begin{equation}\label{final3}
\Delta E^\text{F+Ph}_\text{Cas.}\sim \frac{23S\alpha}{16}\pi^{-7/2}m^{5/2}a^{-1/2}\ln(am)e^{-2am}.
\end{equation}
The pressure on the plates related to this term is
\begin{equation}\label{final3.1}
\Delta P^\text{F+Ph}_\text{Cas.}\sim \frac{23\alpha}{8}\pi^{-7/2}m^3\sqrt{m/a}\ln(am)e^{-2am},
\end{equation}
which clearly shows the exponentially damping structure. In figures \ref{fig1} and \ref{fig2} we compare our result with the leading terms of the Casimir energy for electromagnetic and fermion fields, respectively. Fig. \ref{fig1} shows that the computed correction is negligible even in very small separations. In Fig. \ref{fig2} we see that the impact of this correction increases in large separations.
\section{Conclusions}
 We have calculated one-loop radiative correction to the Casimir energy due to photonic and fermionic counterterms within the renormalized perturbation theory for QED theory. The topology considered here is two perfectly conducting parallel plates in (3+1) dimensions. We have used Dirichlet boundary condition for Electromagnetic field and MIT bag boundary condition for electron. To control ultraviolet divergences we have used dimensional regularization and a photon mass $\mu$ also is used to control infrared divergences. It is found that photonic loop does not have any contribution up to order $\alpha$. The force per unit area related to fermionic and photonic loops, up to this order, at large distances have been obtained as $\Delta P_\text{F}^\text{Cas.}\sim  -\frac{23\alpha}{8}\pi^{-7/2}m^3\sqrt{ma}\ln(am)e^{-2am}$. We illustrate our result in Figs. \ref{fig1} and  \ref{fig2} and compare it with the related leading Casimir energy of electromagnetic and fermion field.

\vskip20pt\noindent {\large {\bf
Acknowledgements}}\vskip5pt\noindent

It is a great pleasure for us to acknowledge the useful discussion and comments
of S.M. Fazeli and M.M. Ettefaghi. This research was supported by the office of research of the
University of Qom.

\section *{Appendix: Calculation Of The Branch-Cut Terms}
In this Appendix we calculate two types of branch-cut terms which appear in Eq. (\ref{1212}). Regardless of some constants, this equation is of the form
\begin{eqnarray}
\sum_{n=1} \sqrt{n^2+b^2}\left( C+ \ln\sqrt{n^2+b^2}\right)-\int_{0}^\infty dx \sqrt{x^2+b^2}\left(  C +\ln\sqrt{x^{2}+b^2}\right)
\end{eqnarray}

In the APSF
\begin{equation}\label{e21:vacc-pol}
\sum_{n=1}^\infty f(n)=-\frac{f(0)}{2}+\int_0^\infty f(x) dx +i\int_0^\infty\frac{dt}{e^{2\pi t}-1}[f(it)-f(-it)]\nonumber,
\end{equation}
Assuming $f(x)= \sqrt{x^2+b^2}\left(  C +\ln\sqrt{x^{2}+b^2}\right)$ we can write
\begin{eqnarray}\label{A.1}
f(it)-f(-it)&=&C(\sqrt{b^2+(it)^2}- \sqrt {b^2+(-it)^2})\nonumber\\
&&\qquad+\left(\sqrt{b^2+(it)^2}\ln\sqrt{b^2+(it)^2}-\sqrt{b^2+(-it)^2}\ln\sqrt{b^2+(-it)^2}\right),
\end{eqnarray}
Choosing, $b=|b|e^{i\theta_{b}}$, $ t=|t|e^{i\theta_t} $ , we have for the first term
\begin{eqnarray}\label{A.2}
\sqrt{b^2+(it)^2}-\sqrt{b^2+(-it)^2}&=&\sqrt{|b|e^{i2\theta_{b}}+e^{i\pi}|t|^2e^{i 2 \theta_t}}-\sqrt{|b|e^{i2\theta_{b}}+e^{-i\pi}|t|^2e^{i 2 \theta_t}}\nonumber \\
&=&\bigg(e^{\frac{i\pi}{2}}e^{i\theta_t}-e^{\frac{-i\pi}{2}}e^{i\theta_t}\bigg)\sqrt{|b|e^{i(2\theta_{b}+\pi-2\theta_t)}+|t|^2}\nonumber \\
&=&2i\sin\left(\frac{\pi}{2}\right)\sqrt{t^2-{b^2}}
\end{eqnarray}
where one should note that $ e^{i(2\theta_{b}+\pi-2\theta_t)}=-1 $ and we assume $t>|b|$. For $t<|b|$ this term is exactly zero. Similarly for second term, we have
\begin{eqnarray}\label{A.4}
&&\hspace{-1cm}\sqrt{b^2+(it)^2}\ln\sqrt{b^2+(it)^2}-\sqrt{b^2+(-it)^2}\ln\sqrt{b^2+(-it)^2}\nonumber\\
&=&\sqrt{b^2+e^{i\pi}t^2}\ln\sqrt{{b}^2+e^{i\pi}t^2}-\sqrt{b^2+e^{-i\pi}t^2}\ln\sqrt{{b}^2+e^{-i\pi}t^2}\nonumber\\
&=&\sqrt{b^2+e^{i\pi}t^2}\ln(e^{i\pi/2}\sqrt{e^{-i\pi}b^2+t^2)})-\sqrt{b^2+e^{-i\pi}t^2}\ln(e^{-i\pi/2}\sqrt{{b}e^{i\pi}+t^2})\nonumber \\
&=&i\frac{\pi}{2}\left[\sqrt{b^2+e^{i\pi}t^2}+\sqrt{b^2+e^{-i\pi}t^2}\right]+\left[\sqrt{b^2+e^{i\pi}t^2}-\sqrt{b^2+e^{-i\pi}t^2}\right]\ln\sqrt{t^2-{b}^2}.
\end{eqnarray}
Now, the first of the last line is similar to similar to \ref{A.2} but with plus sign between its terms, so we get
\begin{equation}\label{A.5}
\sqrt{b^2+e^{i\pi}t^2}+\sqrt{b^2+e^{-i\pi}t^2}=2\sqrt{t^2-{b}^2}\cos\frac{\pi}{2}=0.
\end{equation}
For the second term, using Eq.\eqref{A.2} and Eq.\eqref{A.5}, we have for $t<|b|$
\begin{eqnarray}
\sqrt{b^2+e^{i\pi}t^2}\ln({b}^2+e^{i\pi}t^2)-\sqrt{b^2+e^{-i\pi}t^2}\ln({b}^2+e^{-i\pi}t^2)=2i\sqrt{t^2-{b}}\ln\sqrt{t^2-{b}^2}.
\end{eqnarray}
For $t<|b|$ this term is exactly zero. Therefore, our final result derived as follows:
\begin{equation}
i\int_0^\infty\frac{dt}{e^{2\pi t}-1}[f(it)-f(-it)]=2\int_{b}^\infty
\frac{dt}{e^{2\pi t}-1} \sqrt{t^2-b^2}\left(C- \ln\sqrt{t^2-{b}^2}\right).
\end{equation}

\end{document}